\documentclass[sigconf, nonacm]{acmart}

\usepackage{listings}
\usepackage{xcolor}
\usepackage{colortbl}
\usepackage{pifont}
\usepackage{booktabs}
\usepackage{tikz}
\usetikzlibrary{positioning, arrows.meta, shapes.geometric, fit}
\usepackage{pgfplots}
\pgfplotsset{compat=1.18}
\usepackage{balance}
\usepackage{placeins}
\usepackage{enumitem}
\usepackage{xspace}
\newcommand{\cmark}{\ding{51}}
\newcommand{\xmark}{\ding{55}}
\newcommand{\tool}{\textsc{Agent Audit}\xspace}

\usepackage{xcolor}
\makeatletter
\@ifundefinedcolor{uscred}{\definecolor{uscred}{HTML}{990000}}{}
\makeatother

\usepackage[normalem]{ulem}

\newif\ifshowcomments
\showcommentstrue

\usepackage[table]{xcolor}
\definecolor{lightgray}{gray}{0.95}
\definecolor{lightgreen}{RGB}{220,245,220}
\definecolor{lightblue}{RGB}{230,240,255}

\setcopyright{none}
\acmConference[CAIS 2026]{ACM Conference on AI and Agentic Systems}{May 26--29, 2026}{San Jose, CA, USA}
\acmDOI{}
\acmISBN{}

\begin{document}

\title{\tool{}: A Security Analysis System for LLM Agent Applications}
\thanks{Source code: \url{https://github.com/HeadyZhang/agent-audit} 

Demo video: \url{https://youtu.be/NvRm_14DtbY}}

\author{Haiyue Zhang}
\email{haiyuez@usc.edu}
\affiliation{%
  \institution{University of Southern California}
  \city{Los Angeles}
  \state{California}
  \country{USA}
}

\author{Yi Nian}
\email{yinian@usc.edu}
\affiliation{%
  \institution{University of Southern California}
  \city{Los Angeles}
  \state{California}
  \country{USA}
}

\author{Yue Zhao}
\email{yue.z@usc.edu}
\affiliation{%
  \institution{University of Southern California}
  \city{Los Angeles}
  \state{California}
  \country{USA}
}



\begin{abstract}
What should a developer inspect before deploying an LLM agent: the model, the tool code, the deployment configuration, or all three? 
In practice, many security failures in agent systems arise not from model weights alone, but from the surrounding software stack: tool functions that pass untrusted inputs to dangerous operations, exposed credentials in deployment artifacts, and over-privileged Model Context Protocol (MCP) configurations.

We present \tool{}, a security analysis system for LLM agent applications. \tool{} analyzes Python agent code and deployment artifacts through an agent-aware pipeline that combines dataflow analysis, credential detection, structured configuration parsing, and privilege-risk checks. 
The system reports findings in terminal, JSON, and SARIF formats, enabling direct integration with local development workflows and CI/CD pipelines.
On a benchmark of 22 samples with 42 annotated vulnerabilities, \tool{} detects 40 vulnerabilities with 6 false positives, substantially improving recall over common SAST baselines while maintaining sub-second scan times. \tool{} is open source and installable via \texttt{pip}, making security auditing accessible for agent systems.

In the live demonstration, attendees scan vulnerable agent repositories and observe how \tool{} identifies security risks in tool functions, prompts, and more. Findings are linked to source locations and configuration paths, and can be exported into VS Code and GitHub Code Scanning for interactive inspection.

\end{abstract}

\maketitle

\section{Introduction}

How should developers audit the security of an LLM agent before deployment?
In practice, many failures in agent systems do not arise from model weights alone, but from the surrounding software stack: tool functions that route untrusted inputs to dangerous operations, prompts assembled from unsanitized user content, and deployment artifacts that grant unsafe permissions or expose credentials.
As LLM-based agent applications built with frameworks such as LangChain~\cite{langchain} and CrewAI~\cite{crewai} move into real workflows, these software-layer attack surfaces are becoming increasingly important to inspect.

This setting differs from conventional application security in two ways.
First, agent tool code often sits at a boundary where user input, model output, and external system calls interact.
Second, deployment artifacts such as Model Context Protocol (MCP) configurations~\cite{mcp-spec} introduce additional risks through over-privileged filesystem access, untrusted third-party servers, and embedded secrets. Recent work has demonstrated that indirect prompt injection can compromise LLM-integrated applications by manipulating retrieved data~\cite{greshake2023indirect}, and that tool-augmented agents are vulnerable to injection attacks that hijack tool execution~\cite{zhan2024injecagent, liu2024formalizing}. Disclosed CVEs in popular agent frameworks, including CVE-2023-29374~\cite{cve-2023-29374} and CVE-2023-36258~\cite{cve-2023-36258}, further illustrate that these risks are not hypothetical. Consider the following example.

\begin{lstlisting}[caption={Security risks across agent code and configuration.},label=lst:motivating]
@tool
def analyze_data(expr: str):
    # V1: code execution
    return eval(expr)                  

def build_prompt(user_input: str):
    # V2: prompt injection
    return f"You are a data analyst. {user_input}"   

# MCP configuration
{
  "args": ["@untrusted-org/data-mcp-server", "--allow-write", "/"]  # V3: unsafe MCP configuration
}
\end{lstlisting}

This example illustrates three security surfaces that commonly arise in LLM agent systems.
First, \texttt{@tool}-decorated functions form execution boundaries where user input or model-generated content may reach dangerous operations such as code execution.
Second, prompt-construction code can expose injection surfaces when untrusted input is embedded directly into system instructions.
Third, deployment artifacts such as MCP configurations can introduce additional risks through untrusted third-party servers or over-broad filesystem permissions.

Off-the-shelf Bandit~\cite{bandit} and Semgrep~\cite{semgrep} detect only part of the risk in this example, such as the unsafe \texttt{eval()} call (V1).
They typically do not reason about prompt-construction risks (V2) or MCP-specific configuration semantics (V3).
In contrast, \tool{} analyzes both agent code and configuration artifacts and reports all three issues in this example.

\noindent\textbf{Why existing analyzers are insufficient.}
As shown above, general-purpose SAST tools are effective for many standard code patterns, but agent software introduces structures and threat assumptions that are only partially covered by existing rule sets.\vspace{-1pt}
Examples include tool-boundary reasoning for \texttt{@tool}-decorated functions, prompt-injection surfaces created during prompt assembly, and structured analysis of MCP artifacts.
Table~\ref{tab:capability} summarizes the design goals that distinguish \tool{} from off-the-shelf baselines.\vspace{-1pt}
These goals are also aligned with the OWASP Agentic Security Initiative Top~10~\cite{owasp-agentic-2025}, which provides a recent taxonomy of agent-specific security risks.
To address these gaps, we present \tool{}, a security analysis system for LLM agent applications.
\tool{} analyzes Python agent code and deployment artifacts through a multi-scanner pipeline that combines AST-based dataflow analysis, credential detection, structured configuration parsing, and privilege-risk checks.\vspace{-1pt}
The system produces findings in terminal, JSON, Markdown, and SARIF formats, making it suitable for both local use and CI/CD workflows.

\begin{table}[t]
\caption{Security analysis capability comparison for LLM agent systems. \tool{} provides explicit support for agent-specific analysis that is only partially supported by existing SAST tools.}
\vspace{-5pt}
\label{tab:capability}
\centering
\footnotesize
\scalebox{0.9}{
\begin{tabular}{lcc|c}
\toprule
\rowcolor{lightgray}
\textbf{Capability} & \textbf{Bandit~\cite{bandit}} & \textbf{Semgrep~\cite{semgrep}} & \textbf{\tool{}} \\
\midrule
Agent-tool context awareness      & \xmark & Partial & \cellcolor{lightgreen}\cmark \\
Structured MCP artifact analysis  & \xmark & Partial & \cellcolor{lightgreen}\cmark \\
Prompt-construction risk detection & \xmark & Partial & \cellcolor{lightgreen}\cmark \\
Agent-aware confidence calibration & \xmark & \xmark & \cellcolor{lightgreen}\cmark \\
Agent-specific rule coverage       & \xmark & Partial & \cellcolor{lightgreen}\cmark \\
OWASP ASI-aligned reporting        & \xmark & \xmark & \cellcolor{lightgreen}\cmark \\
\bottomrule
\end{tabular}
}
\end{table}

\noindent
\textbf{Demo plan.}
In the live demonstration (\S\ref{sec:demo}), attendees
scan real-world agent repositories and observe how \tool{}
identifies security risks across tool functions, prompt
construction paths, and MCP deployment configurations.
Findings are linked to source locations and configuration
paths, exportable to VS~Code and GitHub Code Scanning for
interactive triage. \vspace{-1pt}We also demonstrate the \texttt{inspect} subcommand, which
connects to a running MCP server to detect tool-description
poisoning and cross-server shadowing without executing
any tools.\vspace{-1pt}

\noindent\textbf{Contributions.}
We summarize our key contributions as below:
\begin{itemize}[leftmargin=*,nosep]
  \item \textbf{Novelty}. We present \tool{}, a security analysis system for LLM agent applications that analyzes Python agent code and deployment artifacts, including MCP configurations.
  \item \textbf{New benchmark and effectiveness.}
  We introduce Agent-Vuln-Bench (AVB), a new benchmark of
  22~agent-security samples with 42~annotated vulnerabilities
  spanning code execution, credential leakage, and MCP
  configuration risks.
  We evaluate \tool{} against Bandit and Semgrep on AVB,
  showing a 4.0$\times$ recall advantage while maintaining
  sub-second scan times.
  \item \textbf{Accessibility}.
  We demonstrate an open-source implementation with \texttt{pip} installation, SARIF export, and IDE/CI integration, enabling security auditing for diverse agentic workflows.
\end{itemize}

\vspace{-8pt}
\section{System Architecture}

\subsection{Overview}

\begin{figure}[h!]
\centering
\resizebox{\columnwidth}{!}{%
\begin{tikzpicture}[
    every node/.style={font=\scriptsize, align=center},
    inputbox/.style={draw, rounded corners=2pt, fill=blue!8,
                     minimum height=0.45cm, minimum width=1.8cm},
    scanbox/.style={draw, rounded corners=2pt, fill=blue!18,
                    minimum height=0.65cm, minimum width=1.8cm},
    enginebox/.style={draw, rounded corners=2pt, fill=gray!18,
                      minimum height=0.5cm, text width=6.5cm},
    outputbox/.style={draw, rounded corners=2pt, fill=green!12,
                      minimum height=0.4cm, minimum width=1.3cm},
    arr/.style={-{Stealth[length=2pt]}, gray!70},
    layerlabel/.style={font=\scriptsize\itshape, text=gray!60},
]

\node[inputbox] at (0, 0)    (pyfiles)  {Python Files};
\node[inputbox] at (2.5, 0)  (cfgfiles) {Config Files\\(JSON/YAML)};
\node[inputbox] at (5.0, 0)  (allfiles) {All Files};

\node[scanbox] at (0, -1.3)    (pyscan)  {\textbf{Python}\\AST + Taint};
\node[scanbox] at (2.2, -1.3)  (secscan) {\textbf{Secret}\\Regex + Semantic};
\node[scanbox] at (4.4, -1.3)  (mcpscan) {\textbf{MCP Config}\\JSON/YAML};
\node[scanbox] at (6.6, -1.3)  (privscan){\textbf{Privilege}\\AST + Regex};

\draw[arr] (pyfiles.south)  -- (pyscan.north);
\draw[arr] (pyfiles.south)  -- (secscan.north west);
\draw[arr] (cfgfiles.south) -- (secscan.north);
\draw[arr] (cfgfiles.south) -- (mcpscan.north);
\draw[arr] (allfiles.south) -- (mcpscan.north east);
\draw[arr] (allfiles.south) -- (privscan.north);

\node[enginebox] at (3.3, -2.6) (engine)
  {\textbf{RuleEngine}: 57 Rules
   $\rightarrow$ Confidence Tiering $\rightarrow$ Dedup};

\draw[arr] (pyscan.south)   -- ++(0,-0.3) -| ([xshift=-2cm]engine.north);
\draw[arr] (secscan.south)  -- ++(0,-0.2) -| ([xshift=-0.7cm]engine.north);
\draw[arr] (mcpscan.south)  -- ++(0,-0.2) -| ([xshift=0.7cm]engine.north);
\draw[arr] (privscan.south) -- ++(0,-0.3) -| ([xshift=2cm]engine.north);

\node[outputbox] at (1.0, -3.7) (oterm)  {Terminal};
\node[outputbox] at (2.8, -3.7) (ojson)  {JSON};
\node[outputbox] at (4.5, -3.7) (osarif) {SARIF};
\node[outputbox] at (6.2, -3.7) (omd)    {Markdown};

\draw[arr] ([xshift=-1.5cm]engine.south) -- (oterm.north);
\draw[arr] ([xshift=-0.5cm]engine.south) -- (ojson.north);
\draw[arr] ([xshift=0.5cm]engine.south)  -- (osarif.north);
\draw[arr] ([xshift=1.5cm]engine.south)  -- (omd.north);

\node[layerlabel] at (-1.3, 0)    {Input};
\node[layerlabel] at (-1.3,-1.3)  {Scanners};
\node[layerlabel] at (-1.3,-2.6)  {Engine};
\node[layerlabel] at (-1.3,-3.7)  {Output};

\end{tikzpicture}%
}
\caption{System architecture of \tool{}. Four specialized scanners feed into a unified rule engine with confidence-based tiering.}
\label{fig:arch}
\end{figure}
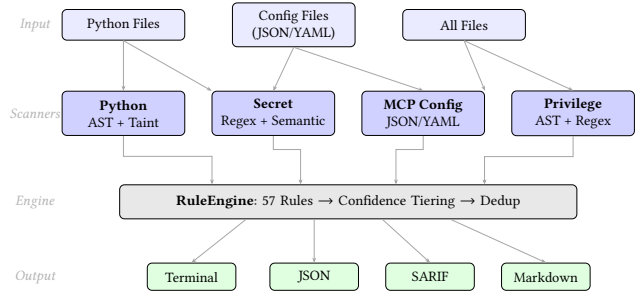

\begin{figure}[h!]
\centering
\resizebox{0.95\columnwidth}{!}{%
\begin{tikzpicture}
\begin{axis}[
    xbar,
    bar width=6pt,
    width=\columnwidth,
    height=5.5cm,
    xlabel={Number of rules},
    xlabel style={font=\scriptsize},
    xmin=0, xmax=13,
    xtick={0,2,4,6,8,10,12},
    xticklabel style={font=\tiny},
    ytick=data,
    yticklabels={
        {ASI-10: Rogue Agents},
        {ASI-09: Trust Exploit.},
        {ASI-08: Cascading},
        {ASI-07: Insecure Comm.},
        {ASI-06: Memory Poison.},
        {ASI-05: Code Execution},
        {ASI-04: Supply Chain},
        {ASI-03: Identity/Priv.},
        {ASI-02: Tool Misuse},
        {ASI-01: Goal Hijack.}
    },
    yticklabel style={font=\tiny},
    y dir=reverse,
    nodes near coords,
    nodes near coords style={font=\tiny, anchor=west},
    every node near coord/.append style={xshift=1pt},
    enlarge y limits=0.08,
    axis line style={gray},
    tick style={gray},
]
\addplot[fill=blue!65, draw=blue!80] coordinates {
    (6,1) (9,2) (9,3) (10,4) (3,5) (2,6) (1,7) (3,8) (7,9) (3,10)
};
\end{axis}
\end{tikzpicture}%
}
\caption{Distribution of \tool{}'s 57 detection rules
across the 10~OWASP Agentic Security Initiative
categories. Four additional cross-cutting rules
(not shown) cover credentials, MCP supply chain, and
privilege escalation.}
\label{fig:owasp}
\end{figure}
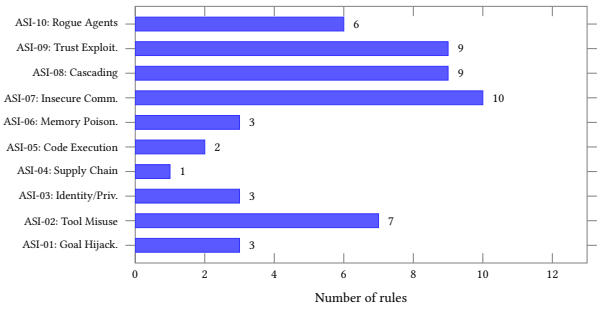

\tool{} employs a multi-scanner pipeline architecture (Figure~\ref{fig:arch}).
Input files are dispatched by type to four specialized scanners that run in parallel.
Raw findings from all scanners are funneled into a unified \emph{RuleEngine} that performs rule ID mapping (73~pattern types $\to$ 57~rules), confidence tiering, and cross-scanner deduplication.
The system outputs findings in four formats: a Rich-based terminal display, JSON, SARIF~\cite{sarif} for CI/CD integration, and Markdown.
\tool{} comprises 22{,}009 lines of Python across 58~source files, with 1{,}323~unit tests.
The 57~rules span all 10~OWASP Agentic Security Initiative
categories (Figure~\ref{fig:owasp}), with the heaviest
coverage on supply chain risks (ASI-04, 10~rules),
tool misuse (ASI-02, 9~rules), and identity/privilege
management (ASI-03, 9~rules).
Four additional cross-cutting rules cover credentials,
MCP supply chain, and privilege escalation.

\vspace{-8pt}
\subsection{Agent-Aware Code Analysis}

\paragraph{Tool Boundary Detection.}
A key design principle is distinguishing \texttt{@tool}-decorated functions from regular code.
\tool{} recognizes 12~tool decorator patterns across LangChain, CrewAI, and custom frameworks (e.g., \texttt{@tool}, \texttt{@langchain.tools.tool}, \texttt{BaseTool} subclasses).
Findings within tool boundaries receive a base confidence of 0.90, versus 0.55 for ordinary functions, reflecting the elevated risk when tool functions process LLM-generated input.
\vspace{-3pt}
\paragraph{Intra-procedural Taint Analysis.}
The PythonScanner implements a four-stage taint pipeline:
(1)~\emph{Source classification}---function parameters, \texttt{chain.invoke()} return values, and \texttt{request.json()} calls are marked as taint sources;
(2)~\emph{Data flow graph construction} via AST traversal;
(3)~\emph{Sanitization detection}---calls to \texttt{shlex.quote()}, \texttt{isinstance()}, or parameterized query binding reduce confidence by $\times$0.20;
(4)~\emph{Sink reachability}---BFS determines whether tainted data reaches dangerous sinks (\texttt{eval()}, \texttt{subprocess.run()}, \texttt{cursor.execute()}).
A dedicated DangerousOperationAnalyzer further reduces false positives for AGENT-034 (tool input validation) by verifying that string parameters in \texttt{@tool} functions actually flow to dangerous operations before flagging.
\vspace{-3pt}
\paragraph{Prompt Injection Surface Detection.}
\tool{} detects user input interpolated into system prompts via f-strings, \texttt{.format()}, string concatenation, and augmented assignment (\texttt{+=}).
Pattern deduplication merges AGENT-010 (generic prompt injection) and AGENT-027 (LangChain-specific) findings on the same line.

\vspace{-5pt}
\vspace{-8pt}
\subsection{Credential and Configuration Analysis}

\paragraph{Credential Detection.}
The SecretScanner uses a three-stage pipeline: (1)~pattern matching with 40+ regular expressions covering API key prefixes, connection strings, and JWT tokens; (2)~semantic analysis using Shannon entropy (rejecting low-entropy placeholders) and framework context detection (suppressing Pydantic \texttt{Field} definitions); (3)~file-path context adjustment (reducing confidence in test fixtures and example files).
\vspace{-8pt}
\paragraph{MCP Configuration Security.}
The MCPConfigScanner parses nine MCP configuration formats (Claude Desktop, VS~Code, Cursor, Windsurf, among others) as structured JSON/YAML---not as source code.
Seven dedicated rules detect overly broad filesystem access (AGENT-029), unverified server sources (AGENT-030), sensitive environment variable exposure (AGENT-031), missing sandboxing (AGENT-032), missing authentication (AGENT-033), schema security (AGENT-040), and excessive server counts (AGENT-042).
Four additional rules target MCP supply-chain attacks: cross-server tool shadowing (AGENT-055), tool description poisoning (AGENT-056), argument injection (AGENT-057), and baseline drift detection (AGENT-054). This analysis is fundamentally beyond the reach of Bandit and Semgrep, which treat JSON files as opaque data.
\vspace{-8pt}
\paragraph{Privilege Escalation Detection.}
The PrivilegeScanner identifies privilege-related risks in
agent deployment configurations using both AST analysis
(Python) and regex matching (JavaScript/shell scripts).
Five rules target: daemon processes
privileges (AGENT-043), \texttt{NOPASSWD} sudoers
entries (AGENT-044), \texttt{CAP\_SYS\_ADMIN}
capabilities (AGENT-045), system credential store access
(AGENT-046), and subprocess execution without sandboxing
(AGENT-047). Confidence is reduced for build scripts and safe commands
(\texttt{git}, \texttt{npm}, \texttt{pip}) to minimize
false positives in CI/CD environments.
\vspace{-10pt}
\subsection{Confidence-Based Tiering}

All findings pass through a four-tier confidence system:
\texttt{BLOCK} ($\geq$0.92) for high-confidence vulnerabilities requiring immediate action,
\texttt{WARN} ($\geq$0.60) for probable issues,
\texttt{INFO} ($\geq$0.30) for informational findings, and
\texttt{SUPPRESSED} ($<$0.30) for likely false positives.
Over 20~false-positive reduction mechanisms: tool boundary boosting, framework path suppression, and test context detection modulate confidence scores.
On synthetic ground-truth fixtures (Layer~1 evaluation), this system achieves 98.51\% precision at 100\% recall.

\FloatBarrier
\vspace{-5pt}
\section{Evaluation}

\subsection{Agent-Vuln-Bench (AVB)}

We construct Agent-Vuln-Bench (AVB), an SWE-bench-style benchmark for evaluating AI agent security scanners.
AVB contains 22~samples organized into three vulnerability sets---injection/RCE (Set~A, 19~vulnerabilities), MCP/components (Set~B, 9), and data/authentication (Set~C, 14)---with 42~expert-annotated vulnerabilities serving as the oracle.
Samples are drawn from CVE reproductions, real-world agent vulnerability patterns, and MCP configuration attacks.
The KNOWN subset includes reproductions of critical CVEs
such as LangChain's \texttt{LLMMath\-Chain}
\texttt{eval()} injection~\cite{cve-2023-29374} and
\texttt{Python\-REPL\-Tool} remote code
execution~\cite{cve-2023-36258}.
The WILD subset captures vulnerability patterns discovered
in production agent code, including calculator tools with
unsandboxed \texttt{eval()}, web-fetcher tools with SSRF
via user-controlled URLs, and agent self-modification through
dynamic \texttt{importlib} usage.
Three samples target MCP-specific supply-chain
attacks: tool shadowing across MCP servers, tool description
poisoning, and baseline configuration drift.
Oracle matching uses file-path suffix and line number.
\vspace{-10pt}
\subsection{Detection Results}

Table~\ref{tab:results} presents the main results.
\tool{} achieves 95.24\% recall and 86.96\% precision (F1\,=\,0.909) on the 42-vulnerability AVB benchmark, compared to 23.8\% recall for Semgrep (F1\,=\,0.385) and 29.7\% for Bandit (F1\,=\,0.458).
The 4.0$\times$ recall advantage over Semgrep is driven by three factors:
(1)~MCP configuration vulnerabilities (10/42 oracle entries) where Semgrep has zero coverage;
(2)~credential detection (12/42) where Semgrep detects only 1;
(3)~agent-specific patterns (prompt injection, self-modification, tool poisoning) absent from generic rule sets.
Of the 42~oracle vulnerabilities, 30 are exclusively detected by \tool{} and zero are exclusively detected by Semgrep.

\begin{table}[h!]
\caption{Detection results on AVB (42 vulnerabilities).}
\label{tab:results}
\centering
\footnotesize
\begin{tabular}{lrrr}
\toprule
\textbf{Metric} & \textbf{\tool{}} & \textbf{Semgrep} & \textbf{Bandit} \\
\midrule
True Positives        & 40     & 10     & 11 \\
False Negatives       &  2     & 32     & 26 \\
False Positives       &  6     &  0     &  0 \\
\midrule
Recall                & \textbf{95.24\%} & 23.8\% & 29.7\% \\
Precision             & 86.96\% & \textbf{100.0\%} & \textbf{100.0\%} \\
F1 Score              & \textbf{0.909} & 0.385 & 0.458 \\
\midrule
Exclusive detections  & 30     &  0     &  1 \\
MCP vuln coverage     & 100\%  &  0\%   &  0\% \\
OWASP ASI coverage    & 10/10  & $\sim$1/10 & $\sim$1/10 \\
\bottomrule
\end{tabular}
\end{table}

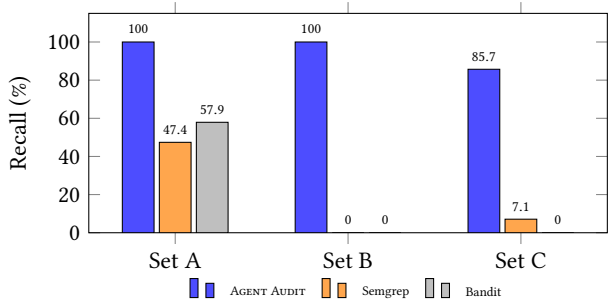
\begin{figure}[h!]
\centering
\begin{tikzpicture}
\begin{axis}[
    ybar,
    bar width=12pt,
    width=\columnwidth,
    height=4.5cm,
    ylabel={Recall (\%)},
    symbolic x coords={Set A, Set B, Set C},
    xtick=data,
    ymin=0, ymax=115,
    ytick={0,20,40,60,80,100},
    legend style={
    at={(0.5,-0.18)},
    anchor=north,
    legend columns=3,
    font=\tiny,
    draw=none,
    inner sep=2pt,
    column sep=4pt,
    },
    enlarge x limits=0.25,
    nodes near coords,
    nodes near coords style={font=\tiny},
    every node near coord/.append style={anchor=south},
]
\addplot[fill=blue!70] coordinates {
    (Set A, 100) (Set B, 100) (Set C, 85.7)
};
\addplot[fill=orange!70] coordinates {
    (Set A, 47.4) (Set B, 0) (Set C, 7.1)
};
\addplot[fill=gray!50] coordinates {
    (Set A, 57.9) (Set B, 0) (Set C, 0)
};
\legend{\tool{}, Semgrep, Bandit}
\end{axis}
\end{tikzpicture}
\caption{Per-set recall comparison on AVB.
\tool{} achieves 100\% recall on injection/RCE and MCP
vulnerability sets, where existing tools have limited or
zero coverage.}
\label{fig:perset}
\end{figure}

Figure~\ref{fig:perset} breaks down detection performance by
vulnerability category.  \tool{} achieves perfect recall on
Set~A (injection/RCE) and Set~B (MCP/components).  The two
false negatives in Set~C are a Markdown-embedded credential
(AGENT-004) and an unimplemented daemon detection rule
(AGENT-043), both in a TypeScript/Markdown noise sample
outside the primary Python scope.
Semgrep and Bandit achieve 100\% precision because they detect so few agent-specific vulnerabilities (10 and 11 TPs respectively, missing 32 and 26 oracle entries).
\tool{}'s six false positives (precision 86.96\%) arise from MCP configuration heuristics flagging safe patterns---an acceptable trade-off given the 4$\times$ recall advantage and 30~exclusive detections.

\vspace{-7pt}
\paragraph{Performance.}
\tool{} scans 22{,}009~lines of Python in 0.87~seconds (25{,}000~lines/sec), matching Bandit's speed (0.90s) while being 6.9$\times$ faster than Semgrep (5.94s) on the same codebase.
Sub-second scan times make \tool{} suitable as a blocking CI check.
Scan time scales linearly with codebase size, with the full test suite (25{,}582~lines) completing in 1.27~seconds.
\vspace{-7pt}
\paragraph{Limitations.}
The current implementation is scoped to intra-procedural taint analysis; inter-procedural data flow across function boundaries is not tracked.
Python is the primary analysis target; TypeScript and JavaScript receive only regex-level scanning.
\tool{} detects code-level vulnerability \emph{patterns} but does not execute or simulate runtime prompt injection payloads.
Confidence thresholds are calibrated empirically rather than derived from a formal model.

\vspace{-15pt}
\section{Demonstration}
\label{sec:demo}

\tool{} is installed via \texttt{pip install agent-audit}
and requires no project-specific configuration.
A single command, \texttt{agent-audit scan ./project},
analyzes all Python files and recognized configuration
artifacts (MCP JSON/YAML, credential files) in the target
directory.
Findings are displayed in a Rich-formatted terminal grouped
by severity tier (\texttt{BLOCK}/\texttt{WARN}/\texttt{INFO}),
with inline source context and remediation guidance.
For CI/CD workflows, \texttt{-{}-format sarif} produces
output compatible with GitHub Code Scanning and VS~Code's
SARIF Viewer.

In the live demonstration, we show \tool{} applied to
three types of real-world agent projects:
\begin{enumerate}[leftmargin=*,nosep,label=(\arabic*)]
  \item \emph{LangChain agent applications}:
  scanning \texttt{@tool}-decorated functions to detect SQL injection, unsandboxed \texttt{eval()}, and prompt injection paths;
  \item \emph{MCP server deployments}:
  scanning Claude Desktop and Cursor configurations for overly broad permissions, unverified sources, and exposed credentials;
  \item \emph{Multi-agent orchestration}:
  scanning a CrewAI project for missing inter-agent authentication and unbounded iteration risks.
\end{enumerate}
For each scenario, attendees observe how adding or removing
a single vulnerability changes the scan output, illustrating
the precision of individual detection rules.
Attendees are invited to install \tool{} on their own
devices and scan their own agent projects during the
session.

We also demonstrate the \texttt{inspect}
subcommand, which connects to a running MCP server via
the stdio transport, enumerates registered tools, and
analyzes their descriptions for instruction overrides,
data exfiltration URLs, and cross-tool manipulation
patterns.
The command detects tool shadowing by comparing tool names
across servers using Levenshtein distance---all without
executing any tools.


\tool{} integrates into CI/CD pipelines via a GitHub
Actions workflow:
\begin{lstlisting}[language={},basicstyle=\ttfamily\scriptsize,
  numbers=none,frame=single,aboveskip=0.5em,belowskip=0.3em,
  xleftmargin=0.5em,framexleftmargin=0.5em]
- uses: HeadyZhang/agent-audit@v1
  with:
    path: '.'
    fail-on: 'high'
    upload-sarif: 'true'
\end{lstlisting}
The \texttt{fail-on} parameter sets the severity threshold
for CI failure, and \texttt{upload-sarif} pushes results to
the GitHub Security tab.
Scanning behavior is customizable via
\texttt{.agent-audit.yaml}, supporting path exclusions,
per-rule suppression, and baseline-relative scanning
(\texttt{-{}-baseline}) for incremental adoption without
alert fatigue.
\footnote{Source code: \url{https://github.com/HeadyZhang/agent-audit}.
\\Demo video: \url{https://youtu.be/NvRm_14DtbY}}

\noindent
\textbf{Artifact availability.}
\tool{} is open-source (MIT license) and available at
\url{https://github.com/HeadyZhang/agent-audit}
and on PyPI via \texttt{pip install agent-audit}.
Example vulnerable agent projects and reproduction
instructions are provided.

\vspace{-8pt}
\section{Related Work}

\paragraph{General-purpose static analysis.}
Bandit~\cite{bandit} performs AST pattern matching for Python security issues; Semgrep~\cite{semgrep} supports generic pattern matching across 30+ languages; CodeQL~\cite{codeql} provides inter-procedural taint analysis via a query language.
None of these tools model agent-specific concepts such as \texttt{@tool} boundaries, MCP configuration semantics, or LLM output as a taint source.
\vspace{-3pt}
\paragraph{Runtime AI security.}
NeMo Guardrails~\cite{nemo-guardrails} and Rebuff~\cite{rebuff} detect prompt injection at runtime; garak~\cite{garak} performs dynamic LLM vulnerability scanning; InjecAgent~\cite{zhan2024injecagent} benchmarks indirect prompt injection in tool-integrated agents.
These tools address runtime threats and are complementary to \tool{}'s static code-level analysis.
\vspace{-3pt}
\paragraph{MCP security tools.}
MCP Checkpoint (aira-security) provides runtime request filtering for deployed MCP servers. \tool{} operates at a different layer, statically analyzing MCP \emph{configurations} before deployment to detect supply-chain risks and credential exposure that runtime filters cannot address.
The two approaches are complementary---\tool{} prevents
insecure configurations from reaching production, while
runtime filters enforce per-request policies on deployed
servers.
\vspace{-3pt}
\paragraph{Standards and frameworks.}
The OWASP Agentic Security Initiative Top~10~\cite{owasp-agentic-2025} provides the first systematic threat taxonomy for AI agent applications.
\tool{} is, to our knowledge, the first static analysis tool to map its rule set to all 10~ASI categories and enforce them at the code level.

\vspace{-8pt}
\section{Conclusion}

Security risks in LLM agent systems arise across multiple layers of the
agent stack, including tool execution, prompt construction, and
deployment configuration.
We introduced \tool{}, a security analysis system that detects these
risks by combining agent-aware code analysis with structured inspection
of MCP configuration artifacts.
The system integrates with practical development workflows through
SARIF output and CI/CD support and is available as an open-source tool.
Future work includes learning-based vulnerability detection and
inference-time monitoring to enable continuous security auditing for
agent systems in production.

\balance
\vspace{-5pt}
\bibliographystyle{ACM-Reference-Format}
\bibliography{references}

\end{document}